\documentclass[conference]{IEEEtran}
\IEEEoverridecommandlockouts
\usepackage{cite}
\usepackage{amsmath,amssymb,amsfonts}
\usepackage{algorithmic}
\usepackage{graphicx}
\usepackage{textcomp}
\usepackage{xcolor}
\usepackage{arydshln}
\usepackage{booktabs}
\usepackage[flushleft]{threeparttable}
\usepackage{hyperref}
\hypersetup{
    colorlinks=true,
    linkcolor=blue,
    filecolor=magenta,      
    urlcolor=cyan,
}
\usepackage{footnote}
\usepackage{etoolbox}
\usepackage[super]{nth}
\appto\TPTnoteSettings{\footnotesize}
\def\BibTeX{{\rm B\kern-.05em{\sc i\kern-.025em b}\kern-.08em
    T\kern-.1667em\lower.7ex\hbox{E}\kern-.125emX}}
\begin{document}

\title{Toward Drug-Target Interaction Prediction via Ensemble Modeling and Transfer Learning}

\makeatletter
\newcommand{\linebreakand}{%
  \end{@IEEEauthorhalign}
  \hfill\mbox{}\par
  \mbox{}\hfill\begin{@IEEEauthorhalign}
}
\makeatother

\author{

\IEEEauthorblockN{Po-Yu Kao}
\IEEEauthorblockA{\textit{Insilico Medicine Taiwan Ltd.} \\
Taipei, Taiwan \\
ken.kao@insilicomedicine.com}
\and
\IEEEauthorblockN{Shu-Min Kao}
\IEEEauthorblockA{\textit{Insilico Medicine Taiwan Ltd.} \\
Taipei, Taiwan \\
allen.kao@insilicomedicine.com}
\linebreakand
\IEEEauthorblockN{Nan-Lan Huang}
\IEEEauthorblockA{\textit{Insilico Medicine Taiwan Ltd.} \\
Taipei, Taiwan \\
shannon.huang@insilicomedicine.com}
\and
\IEEEauthorblockN{Yen-Chu Lin}
\IEEEauthorblockA{\textit{Insilico Medicine Taiwan Ltd.} \\
Taipei, Taiwan \\
jimmy.lin@insilicomedicine.com}

}

\maketitle

\begin{abstract}

Drug-target interaction (DTI) prediction plays a crucial role in drug discovery, and deep learning approaches have achieved state-of-the-art performance in this field.
We introduce an ensemble of deep learning models (EnsembleDLM) for DTI prediction.
EnsembleDLM only uses the sequence information of chemical compounds and proteins, and it aggregates the predictions from multiple deep neural networks.
This approach not only achieves state-of-the-art performance in Davis and KIBA datasets but also reaches cutting-edge performance in the cross-domain applications across different bio-activity types and different protein classes.
We also demonstrate that EnsembleDLM achieves a good performance (Pearson correlation coefficient and concordance index $>$ 0.8) in the new domain while the training set has twice as much data as the test set with transfer learning.

\end{abstract}

\begin{IEEEkeywords}
drug-target interaction prediction, ensemble modeling, transfer learning, convolutional neural network, deep neural network
\end{IEEEkeywords}

\section{Introduction}

Drug-target interaction (DTI) prediction plays an important role in computer-aided drug discovery.
It is an essential step of virtual screening and drug repositioning.
The binding affinity of a drug-target pair is generally expressed in terms of dissociation constant ($K_d$), inhibition constant ($K_i$), or the half-maximal inhibitory concentration ($IC_{50}$).
The lower $K_d$, $K_i$, or ${IC}_{50}$ values, the better binding affinities of drug-target pairs.  
The most commonly used technique to predicting the binding affinity of drug-target pairs is docking simulation, e.g., Glide \cite{friesner2004glide}, MOE-Dock \cite{vilar2008medicinal}, AutoDock Vina \cite{trott2010autodock}.
However, different docking programs may result in different docking scores for the same drug-target pair \cite{wang2016comprehensive}. 
In addition, docking is a computationally expensive and time-consuming process \cite{hassan2017protein}, and it can only be applied to proteins whose 3D structures are determined.
Therefore, it is difficult to be applied on a large scale.
In spite of that, our method only considers the sequence information of the chemical compounds and the proteins, i.e., it does not require the 3D structure information of proteins. 
Hence, the proposed method is suitable for screening a huge number of drug candidates and target proteins, and it can be easily applied to virtual screening and drug repositioning.

Recently, learning-based algorithms have been developed for drug-target interaction prediction over the last decade. 
Pahikkala et al. \cite{pahikkala2015toward} used a Kronecker Regularized Least Squares (KronRLS) method to predict the drug-target interactions based on the chemical structure and protein sequence similarity matrices.
He et al. \cite{he2017simboost} utilized gradient boosting machines called \textit{SimBoost} to predict drug-target binding affinities.  
{\"O}zt{\"u}rk et al. \cite{ozturk2018deepdta} presented a deep learning-based model called DeepDTA that uses the sequence information of drugs and targets to predict their binding affinities.  
Shin et al. \cite{shin2019self} introduced a new molecule representation based on the self-attention mechanism and a new deep-learning model called Molecule Transformer DTI (MT-DTI) to predict the binding affinities of chemical compounds and proteins. 
Zhao et al. \cite{zhao2019attentiondta} proposed a deep learning-based end-to-end model named AttentionDTA that uses attention mechanisms on both drug and target sequences to predict their binding affinity scores. 
Abbasi et al. \cite{abbasi2020deepcda} presented a deep learning-based model called DeepCDA that integrates long short-term memory (LSTM) and convolutional neural network (CNN) to predict drug-target interactions. 
It is noted that the methods mentioned above are only based on the sequence information of chemical compounds and protein. 

In this paper, we propose an ensemble of deep learning models (EnsembleDLM) for drug-target interaction prediction using only sequence information of drug and target. 
Our contribution is twofold. 
First, EnsembleDLM has better intra- and inter-domain drug-target interaction prediction performance compared to the sequence-based state-of-the-art methods. 
Second, we have examined how much data a network needs to achieve a good drug-target interaction prediction performance via transfer learning.

\section{Materials and Methodology}

\subsection{Datasets}

\subsubsection{Davis} 
Davis et al. \cite{davis2011comprehensive} curated 30,056 $K_d$ binding affinity scores from 68 chemical compounds and 72 kinase inhibitors with 442 kinases. 
In the Davis dataset, the lower $K_d$ score indicates the better binding affinity of the drug-target pair.

\subsubsection{KIBA}
Tang et al. \cite{tang2014making} introduced a model-based approach called Kinase Inhibitor Bio-Activity (KIBA) to integrate three large-scale biochemical assays including $IC_{50}, K_i$ and $K_d$ of kinase inhibitors. 
The original KIBA dataset provides a total of 246,088 KIBA scores across 52,498 chemical compounds and 467 kinase targets. 
The lower KIBA score indicates the better binding affinity of the drug-target pair. 
However, we use the modified KIBA dataset curated by He et al. \cite{he2017simboost} in this paper.   
The modified KIBA dataset contains 118,254 drug-target bio-activity scores across 2,111 chemical compounds and 229 kinase targets. 
The higher modified KIBA score indicates the better binding affinity of the drug-target pair. 

\subsubsection{ChEMBL} 
ChEMBL \cite{gaulton2017chembl} is an open large-scale bio-activity dataset. 
In this paper, we curated the interaction of drug-target pairs based on the following criteria in ChEMBL:
\begin{itemize}
\item standard\_type == ``Ki''
\item standard\_value $>$ 0
\item standard\_units == ``nM''
\item standard\_relation == ``=''
\item bao\_format == ``BAO\_0000357'' (meaning that in the assays, only results with single protein format were considered)
\end{itemize}
For a drug-target interaction with multiple measurements, the minimum value was taken.
In addition, identical values were removed due to duplication.
In the end, 137,413 drug-target interactions are curated in our dataset.
We then split the curated interactions based on the protein family classification, and remove 894 interactions of unclassified proteins (UP) from the data.
The details of splitting the data are described in Table \ref{tab:chembl}.
In the ChEMBL-$K_i$ dataset, the lower $K_i$ score indicates the better binding affinity of the drug-target pair.

\begin{table}[htbp!]
\centering
\renewcommand*{\arraystretch}{1.4}
\caption{The details of splitting the ChEMBL data into five different classes based on the protein structures. UP : Unclassified proteins.}
\begin{tabular}{rrr} 
\toprule
Label & Level-1              & Level-2                \\ 
\midrule
EK    & Enzyme               & Kinase                 \\
EP    & Enzyme               & Protease               \\
ER    & Epigenetic regulator & Eraser; Reader; Writer \\
TFNR  & Transcription factor & Nuclear receptor       \\
REST  & \multicolumn{2}{c}{(All interactions) - (EK+EP+ER+TFNR+UP)} \\ 
\bottomrule
\end{tabular}
\label{tab:chembl}
\end{table}

More details of these datasets can be found in Table \ref{tab:datasets}.

\begin{table*}[htbp!]
\centering
\renewcommand*{\arraystretch}{1.4}
\caption{The details of datasets contain the number of unique drugs, the number of unique targets, the number of interactions, bio-activity type, and statistics of SMILES sequence length and FASTA sequence length.}
\resizebox{\textwidth}{!}{%
\begin{tabular}{@{\extracolsep{4pt}}lcccccccccc@{}}
\toprule
& & & & & \multicolumn{3}{c}{SMILES sequence length} & \multicolumn{3}{c}{FASTA sequence length} \\ \cmidrule{6-8}  \cmidrule{9-11}
& \# drugs & \# targets & \# interactions & Bio-activity type & Max. & Min.  & Avg. & Max. & Min. & Avg. \\ 
\midrule
Davis & 68 & 442 & 30,056 & $K_d$ & 92 & 39 & 63 & 2,549 & 244  & 789 \\
KIBA & 2,111 & 229 & 118,254 & Mixture of $K_d, K_i$ and $IC_{50}$  & 590 & 20 & 55 & 4,128 & 215 & 731 \\ 
ChEMBL-EK & 7,330 & 188  & 10,508 & $K_i$ & 417 & 12 & 53 & 4,128 & 270 & 957 \\
ChEMBL-EP & 21,392 & 245 & 34,660 & $K_i$ & 718 & 9 & 70 & 3,391 & 99 & 496 \\
ChEMBL-ER & 1,043 & 53 & 1,534 & $K_i$ & 445 & 12 & 63 & 3,046 & 347 & 1,051 \\
ChEMBL-REST & 47,774 & 1,116 & 86,016 & $K_i$ & 837 & 3 & 54 & 7,078 & 26 & 438 \\ 
ChEMBL-TFNR & 2,404 & 40 & 3,776 & $K_i$ & 451 & 24 & 63 & 984 & 352 & 677 \\ \bottomrule
\end{tabular}
}
\begin{tablenotes}
\item KIBA is curated by He et al. \cite{he2017simboost}.
\end{tablenotes}
\label{tab:datasets}
\end{table*}

\subsection{Network Architecture}

The network architecture of the individual model is shown in Fig. \ref{fig:network}. 
The network comprises two pathways which are a drug encoder, target encoder, and regressor. 
Drug encoder and target encoder respectively take a SMILES sequence and a FASTA sequence as an input and output its corresponding feature vector. 
The drug feature vector is then concatenated with the target feature vector and feed into the regressor.
In the end, the network outputs the predicted binding affinity score of the given drug-target pair. 
More details of drug pathway, target pathway, and regressor can be found in Table \ref{tab:network}.

\begin{figure}[ht]
\centering
\includegraphics[width=0.9\linewidth]{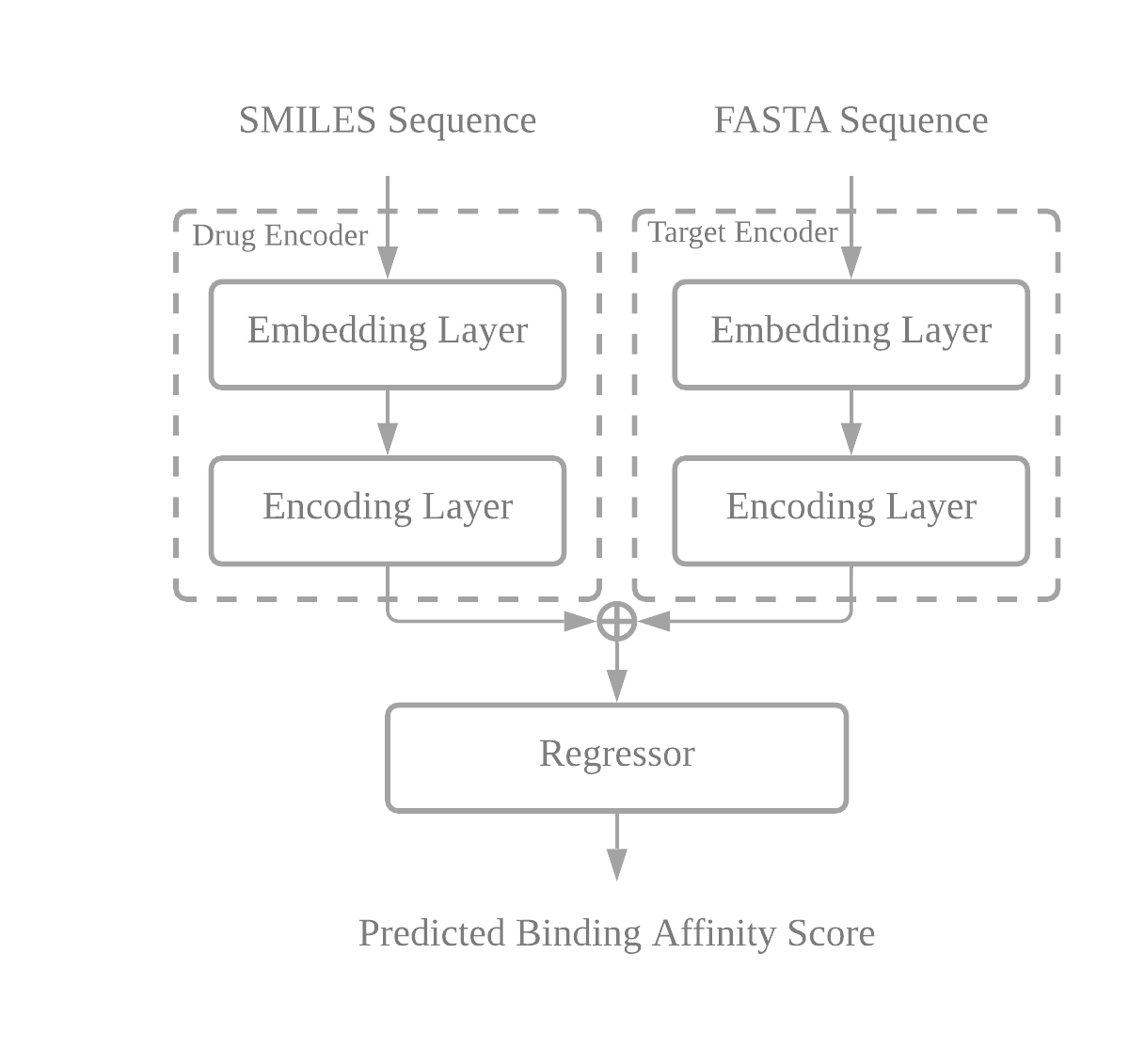}
\caption{The network architecture of individual model.}
\label{fig:network}
\end{figure}

\subsubsection{Drug Encoder} 

The drug encoder contains an embedding layer and an encoding layer.  
Three pre-defined embeddings including Morgan fingerprint \cite{rogers2010extended}, ErG fingerprint \cite{stiefl2006erg}, and Daylight fingerprint \cite{daylight} and one character-based embedding are used in the drug encoder.
The one-hot encoder converts the SMILES characters into a binary matrix for the character-based embedding. 
The encoder layers consist of multiple fully connected layers for the pre-defined embeddings and multiple 1D convolutional layers with one fully connected layer at the end for the character-based embedding.

\subsubsection{Target Encoder}

The target encoder also has an embedding layer and an encoding layer. 
Amino acid composition (AAC) \cite{smith1966amino} and character-based embedding are used to convert the FASTA sequence into a binary vector. 
The one-hot encoder converts the FASTA characters into a binary matrix for the character-based embedding. 
The encoder layers contain multiple fully connected layers for AAC and multiple 1D convolutional layers with one fully connected layer at the end for the character-based embedding.

\subsubsection{Regressor}

The regressor begins with a dropout layer \cite{srivastava2014dropout} with a dropout rate of 0.1 to prevent overfitting followed by multiple fully connected layers.   
It takes the concatenated feature vector from the drug encoder and target encoder as input and outputs the predicted binding affinity score for the corresponding drug-target pair.

\begin{table*}[htbp!]
\small
\renewcommand*{\arraystretch}{1.4}
\caption{The details of network architecture.}
\centering
\resizebox{\textwidth}{!}{%
\begin{threeparttable}
\begin{tabular}{@{\extracolsep{4pt}}cccccccc@{}}
\toprule
& \multicolumn{4}{c}{Drug Pathway} & \multicolumn{2}{c}{Target Pathway} & \\ \cmidrule{2-5}  \cmidrule{6-7}
& Daylight & ErG & Morgan & CNN & AAC & CNN & Regressor \\ 
\midrule
\nth{1} layer & Linear(2048, 1024)  & Linear(315, 1024) & Linear(1024, 1024) & Conv1d(63, 32, 4) & Linear(8420, 1024) & Conv1d(26, 32, 4) & Linear(512, 1024) \\
\nth{2} layer & Linear(1024, 256) & Linear(1024, 256)   & Linear(1024, 256) & Conv1d(32, 64, 4) & Linear(1024, 256) & Conv1d(32, 64, 4) & Linear(1024, 1024) \\
\nth{3} layer & Linear(256, 64) & Linear(256, 64) & Linear(256, 64) & Conv1d(64, 96, 4) & Linear(256, 64) & Conv1d(64, 96, 4) & Linear(1024, 512) \\
\nth{4} layer & Linear(64, 256) & Linear(64, 256) & Linear(64, 256) & Linear(96, 256) & Linear(64, 256) & Linear(96, 256) & Linear(512, 1) \\\bottomrule
\end{tabular}
\begin{tablenotes}
\item Linear(X, Y): Linear layer with X input features, Y output features and bias=True. 
\item Conv1D(X, Y, Z): 1D convolutional layer with X input channels, Y input channels, and Z convolving kernel size.
\end{tablenotes}
\end{threeparttable}
}
\label{tab:network}
\end{table*}

\subsection{Ensemble Modeling} 

Ensembles have been proven to have better performance than any single model \cite{dietterich2000ensemble} in different applications including image classification \cite{he2016deep}, pulmonary tuberculosis diagnosis \cite{lakhani2017deep,alves2019ensemble}, brain tumor segmentation \cite{kamnitsas2017ensembles,kao2018brain,kao2020improving}, solar radiation forecasting \cite{voyant2017machine,guermoui2020comprehensive}, and cancer classification \cite{tan2003ensemble,liu2010ensemble}.
In this paper, we integrate multiple models by taking the arithmetic mean on the predicted binding affinity scores of different models that are defined as 

\[\hat{y}_j = \frac{1}{M}\sum_{i=1}^M \overline{y}_{ij}\] 

where M is the number of models, and $\hat{y}_j$ and $\overline{y}_{ij}$ is the predicted binding scores of ensemble and individual model $i$, respectively, for drug-target pair $j$.

\subsection{Transfer Learning} 

Transfer learning \cite{pan2009survey} is the idea of transferring the domain knowledge from previously learned tasks to a new but related task. 
There are two different methods of transfer learning in neural networks. 
The first one is to use a pre-trained neural network as a feature extractor and concatenate this feature extractor by one or more fully connected layers. 
We only train the concatenated fully connected layers with the new data. 
The second one is to initialize the neural network with pre-trained weights rather than random weights and fine-tune the whole or last few layers with the new data. 
In this paper, we focus on initializing the network with the pre-trained weights and fine-tuning the whole neural network.
Another contribution of this study is to examine how much data does a network needs to achieve a good drug-target interaction prediction performance via transfer learning. 

\subsection{Training and Inference}

\subsubsection*{Training} 

In this study, we use $pK_d$ for Davis and $pK_i$ for ChEMBL to train the models.
\[ pK_d = -log_{10}(\frac{K_d}{10^9}) \quad \text{and} \quad pK_i = -log_{10}(\frac{K_i}{10^9})\]
The larger the $pK_d$ and $pK_i$ score, the better the binding affinity of the drug-target pair. 
In addition, we use the modified KIBA score \cite{he2017simboost} to train the models. 
The larger the modified KIBA score, the better the binding affinity of the drug-target pair. 
During training, models use the Mean Square Error (MSE) as the loss function and adaptive moment estimation (Adam) \cite{kingma2014adam} in the optimization step.
The average training time is approximately 0.23 seconds for one iteration with a batch size of 128 on Intel(R) Xeon(R) Platinum 8260 CPU @ 2.40GHz, NVIDIA GeForce RTX 2080 Ti 11GB GDDR6 and 256G DDR4 RAM. 
The overall pipeline and models are adapted from DeepPurpose \cite{huang2020deeppurpose} with Ax \cite{bakshy2018ae}. 

\subsubsection*{Hyper-parameter Tuning}

Hyper-parameter tuning is an essential step to find a set of optimal hyper-parameters for training a neural network. 
In this paper, we only focus on tuning the hyper-parameters which determine how the network is trained by utilizing Bayesian optimization \cite{snoek2012practical}.
These hyper-parameters include the learning rate searching over $[10^{-6}, 10^{-3}]$, the weight decay searching over $[0, 0.2]$, the number of training epochs searching over $\{100, 120, 150, 200\}$, and the batch sizes searching over $\{128, 256, 512\}$.  
The same set of hyper-parameters is used for both training a network from scratch and fine-tuning a network with pre-trained weights. 

\subsubsection*{Inference} 

In the inference phase, we select the model with the lowest MSE from five-fold cross-validation, and the predicted binding affinity score of a drug-target pair is averaged from seven different models (Daylight-AAC, Morgan-AAC, ErG-AAC, Daylight-CNN, Morgan-CNN, ErG-CNN, and CNN-CNN).

\section{Experiments and Results}

\subsection{Evaluation Metrics}

\subsubsection{Mean squared error} The mean square error (MSE) is used to measure the difference between the predicted binding affinity scores and the ground-truth binding affinity scores, and it is defined as follows:

\[ MSE =  \frac{1}{N}\sum_{j=1}^N (\hat{y}_j-y_j)^2\]

where $N$ is the number of drug-target pairs, and $\hat{y}_j$ and $y_j$ are the predicted and ground-truth binding affinity scores for the drug-target pair $j$, respectively. 
$MSE=0$ corresponds to the perfect model prediction.

\subsubsection{Concordance index} 
The concordance index (C-index) \cite{harrell1996multivariable}, which quantifies the quality of rankings, is used to measure the order of the predictions versus the order of the ground truths. 
The C-index is a value between 1.0 and 0.0.
A value of 1.0 indicates a perfect separation of drug-target pairs with different binding affinity scores and a value of 0.5 indicates no predictive discrimination. 

\subsubsection{Pearson correlation coefficient} 
The Pearson correlation coefficient (PCC) \cite{benesty2009pearson} provides a measure of the strength of linear association between the predicted and ground-truth binding affinity scores.
Given a pair of binding affinity scores $(\hat{Y}, Y)$, their Pearson correlation coefficient $\rho_{\hat{Y},Y}$ is:

\[ \rho_{\hat{Y},Y} = \frac{cov(\hat{Y},Y)}{\sigma_{\hat{Y}}\sigma_{Y}} \]

where $\sigma_{\hat{Y}}$ and $\sigma_{Y}$ are the standard deviations of the predicted and ground-truth binding affinity scores, respectively, and $cov$ is the co-variance.  
The range of the Pearson correlation coefficient is between 1.0 and -1.0. 1.0, -1.0, and 0.0 indicate a strong positive, strong negative, and zero correlation, respectively, between predicted and ground-truth binding affinity scores.

\subsection{Single model vs. ensemble modeling} \label{subsec:single_ensemble}

In this section, we would like to examine the performance of each model with different drug-target encoder combinations and compare their performance with the proposed ensemble modeling approach on the Davis, KIBA, and ChEMBL-REST datasets. 
The proposed ensemble of deep learning models (EnsembleDLM) contains the Daylight-AAC model, Morgan-AAC model, ErG-AAC model, Daylight-CNN model, Morgan-CNN model, ErG-CNN model, and CNN-CNN model.
We first leave 20\% of the data (6,012 interactions for Davis, 23,651 interactions for KIBA, and 17,190 interactions for ChEMBL-REST) out for the test. 
Then, we implement the five-fold cross-validation within the remaining data (24,044 interactions for Davis, 94,603 interactions for KIBA, and 68,759 interactions for ChEMBL-REST).
The remaining data is equally split into five subsets.
Each subset is then taken in turn as a validation set, and the remaining four sets are used to train the model.
Cross-validation is used to prevent over-fitting \cite{ng1997preventing}. 
The performance of each model and proposed EnsembleDLM on the validation set and test set of Davis and KIBA are presented in Table \ref{tab:single_model_davis_kiba} and ChEMBL-REST is presented in Table \ref{tab:single_model_chembl}.

\begin{table*}[htbp!]
\renewcommand*{\arraystretch}{1.4}
\centering
\caption{The comparison of our proposed approach and each individual model on the validation set and test set of Davis and KIBA with 5-fold cross-validation. The individual model is represented by (drug encoder)-(target encoder). The bold number highlights the best performance. The results are rounded down to three decimal places.}
\resizebox{\textwidth}{!}{%
\begin{tabular}{@{\extracolsep{4pt}}ccccccccc@{}}
\toprule
            & \multicolumn{4}{c}{Davis}                            & \multicolumn{4}{c}{KIBA}                                                        \\ \cmidrule{2-5} \cmidrule{6-9} 
            & \multicolumn{2}{c}{Validation} & \multicolumn{2}{c}{Test} & \multicolumn{2}{c}{Validation}    & \multicolumn{2}{c}{Test}                         \\ \cmidrule{2-3} \cmidrule{4-5} \cmidrule{6-7} \cmidrule{8-9}  
Drug-Target & MSE$\pm$std  & C-index$\pm$std & MSE$\pm$std & C-index$\pm$std & MSE$\pm$std  & C-index$\pm$std & MSE$\pm$std & C-index$\pm$std \\ \midrule \midrule
Daylight-AAC                    & 0.256$\pm$0.011 & 0.875$\pm$0.006 & 0.251$\pm$0.003 & 0.874$\pm$0.001 & 0.170$\pm$0.012 & 0.874$\pm$0.002 & 0.168$\pm$0.003 & 0.877$\pm$0.002 \\
Morgan-AAC                      & 0.268$\pm$0.012 & 0.871$\pm$0.006 & 0.256$\pm$0.008 & 0.875$\pm$0.004 & 0.178$\pm$0.004 &  0.875$\pm$0.001 & 0.175$\pm$0.003 & 0.877$\pm$0.003 \\
ErG-AAC                         & 0.261$\pm$0.024 & 0.874$\pm$0.013 & 0.253$\pm$0.006 & 0.879$\pm$0.001 & 0.194$\pm$0.005 & 0.869$\pm$0.002 & 0.188$\pm$0.003 & 0.868$\pm$0.003 \\
Daylight-CNN                    & 0.251$\pm$0.005 & 0.880$\pm$0.006 & 0.249$\pm$0.007 & 0.876$\pm$0.005 & 0.174$\pm$0.005 & 0.876$\pm$0.003 & 0.168$\pm$0.005 & 0.877$\pm$0.003 \\
Morgan-CNN                      & 0.244$\pm$0.012 & 0.881$\pm$0.006 & 0.234$\pm$0.004 & 0.881$\pm$0.005 & 0.171$\pm$0.003 & 0.880$\pm$0.003 & 0.169$\pm$0.004 & 0.881$\pm$0.003 \\
ErG-CNN                         & 0.243$\pm$0.010 & 0.883$\pm$0.006 & 0.242$\pm$0.011 & 0.881$\pm$0.004 & 0.194$\pm$0.006 & 0.869$\pm$0.003 & 0.187$\pm$0.004 & 0.869$\pm$0.004 \\
CNN-CNN                         & 0.250$\pm$0.008 & 0.882$\pm$0.006 & 0.237$\pm$0.003 & 0.883$\pm$0.005 & 0.184$\pm$0.009 & 0.868$\pm$0.005 & 0.181$\pm$0.005 &   0.868$\pm$0.006   \\
\textbf{Proposed approach}      & \textbf{0.202$\pm$0.006} & \textbf{0.907$\pm$0.004} & \textbf{0.204$\pm$0.002} & \textbf{0.905$\pm$0.001} & \textbf{0.138$\pm$0.003} & \textbf{0.895$\pm$0.001} & \textbf{0.138$\pm$0.003} & \textbf{0.896$\pm$0.001} \\ \bottomrule
\end{tabular}
}
\label{tab:single_model_davis_kiba}
\end{table*}

\begin{table*}[htbp!]
\renewcommand*{\arraystretch}{1.4}
\centering
\caption{The comparison of our proposed approach and each individual model on the validation set and test set of ChEMBL-REST with 5-fold cross-validation. The individual model is represented by (drug encoder)-(target encoder). The bold number highlights the best performance. The results are rounded down to three decimal places.}
\begin{tabular}{@{\extracolsep{4pt}}ccccc@{}}
\toprule
            & \multicolumn{4}{c}{ChEMBL-REST}                                                        \\ \cmidrule{2-5}  
            & \multicolumn{2}{c}{Validation} & \multicolumn{2}{c}{Test} \\ \cmidrule{2-3} \cmidrule{4-5}   
Drug-Target & MSE$\pm$std  & C-index$\pm$std & MSE$\pm$std & C-index$\pm$std \\ \midrule \midrule
Daylight-AAC                    & 0.868$\pm$0.060 & 0.820$\pm$0.003 & 0.830$\pm$0.017 & 0.820$\pm$0.003 \\
Morgan-AAC                      & 0.848$\pm$0.040 & 0.822$\pm$0.004 & 0.815$\pm$0.037 & 0.823$\pm$0.003 \\
ErG-AAC                         & 0.937$\pm$0.072 & 0.813$\pm$0.003 & 0.884$\pm$0.028 & 0.812$\pm$0.003 \\
Daylight-CNN                    & 0.905$\pm$0.076 & 0.815$\pm$0.004 & 0.865$\pm$0.024 & 0.815$\pm$0.005\\
Morgan-CNN                      & 0.855$\pm$0.068 & 0.822$\pm$0.002 & 0.823$\pm$0.028 & 0.821$\pm$0.004 \\
ErG-CNN                         & 0.919$\pm$0.032 & 0.814$\pm$0.002 & 0.878$\pm$0.018 & 0.811$\pm$0.003 \\
CNN-CNN                         & 0.925$\pm$0.075 & 0.809$\pm$0.002 & 0.894$\pm$0.019 & 0.809$\pm$0.002 \\
\textbf{Proposed approach}      & \textbf{0.636$\pm$0.035} & \textbf{0.845$\pm$0.002} & \textbf{0.638$\pm$0.008} & \textbf{0.845$\pm$0.001}\\ \bottomrule
\end{tabular}
\label{tab:single_model_chembl}
\end{table*}

\subsection{Compared to the state-of-the-art models}

In this section, we compare the performance of our proposed approach with other state-of-the-art models. 
It is noted that for a fair comparison, we only compare our proposed approach with the sequence-based models, i.e., the inputs to the model are SMILES sequence and FASTA sequence. 
Davis and KIBA are used to compare the performance of different methods. 
We also withhold 20\% of the data for the test and perform five-fold cross-validation on the remaining data.
The comparisons of our approach and the state-of-the-art approaches are shown in Table \ref{tab:davis} for Davis and Table \ref{tab:kiba} for KIBA.

\begin{table}[htbp!]
\renewcommand*{\arraystretch}{1.4}
    \centering
    \caption{The comparison of our approach and the state-of-the-art approaches with 5-fold cross-validation on the Davis dataset. The bold number highlights the best performance. The results are rounded down to three decimal places.}
    \begin{tabular}{ l c c }\toprule
        Models & MSE($\pm$std) & C-index$\pm$std \\\midrule
        KronRLS\cite{pahikkala2015toward} & 0.379 & 0.782$\pm$0.001 \\
        SimiBoost\cite{he2017simboost} & 0.282 & 0.836$\pm$0.001 \\
        DeepDTA\cite{ozturk2018deepdta} & 0.261 & 0.863$\pm$0.002 \\
        MT-DTI\cite{shin2019self} & 0.245 & 0.882$\pm$0.001 \\
        AttentionDTA\cite{zhao2019attentiondta} & 0.216$\pm$0.019 & 0.882$\pm$0.004 \\
        DeepCDA\cite{abbasi2020deepcda} & 0.248 & 0.889$\pm$0.002 \\
        \textbf{Proposed approach} & \textbf{0.202$\pm$0.006} & \textbf{0.907$\pm$0.004}  \\\bottomrule
    \end{tabular}
    
    \label{tab:davis}
\end{table}

\begin{table}[htbp!]
\renewcommand*{\arraystretch}{1.4}
    \centering
    \caption{The comparison of our approach and state-of-the-art approaches with 5-fold cross-validation on the KIBA dataset. The bold number highlights the best performance. The results are rounded down to three decimal places.}
    \begin{tabular}{ l c c }\toprule
        Models & MSE($\pm$std) & C-index$\pm$std \\\midrule
        KronRLS\cite{pahikkala2015toward} & 0.411 & 0.782$\pm$0.001 \\
        SimiBoost\cite{he2017simboost} & 0.222 & 0.836$\pm$0.001\\
        DeepDTA\cite{ozturk2018deepdta} & 0.194 & 0.863$\pm$0.002 \\
        MT-DTI\cite{shin2019self} & 0.152 & 0.882$\pm$0.001 \\
        AttentionDTA\cite{zhao2019attentiondta} & 0.155$\pm$0.003 & 0.882$\pm$0.004 \\
        DeepCDA\cite{abbasi2020deepcda} & 0.176 & 0.889$\pm$0.002 \\
        \textbf{Proposed approach} & \textbf{0.138$\pm$0.003} & \textbf{0.895$\pm$0.001}  \\\bottomrule
    \end{tabular}
    \label{tab:kiba}
\end{table}

\subsection{Cross-domain Drug-Target Interaction Prediction via Fine-Tuning}

In this section, we would like to examine the cross-domain drug-target interaction prediction performance of EnsembleDLM between different datasets including Davis, KIBA, and Ki-ChEMBL.
For different types of models, we select the best one from five-fold cross-validation in the original domain.
We then fine-tune each selected model with the original hyper-parameter set to the new domain using different portions of data, i.e., the Bayesian optimization is not used at this time. 
It is noted that all layers of drug encoder, target encoder, and regressor are fine-tuned.
In the end, the output of our fine-tuned EnsembleDLM is the average predictions of seven fine-tuned models.

\subsubsection{Same protein class with different types of binding affinity scores}
In this experiment, we would like to examine the performance of EnsembleDLM across different binding affinity scores in the same protein class. 
Davis and KIBA are used in this experiment.
Davis consists of $K_d$ interactions of enzyme inhibitions, and KIBA combines $K_d$, $K_i$, and $IC_{50}$ interactions of enzyme inhibitions into a new score.
We hold 20\% of new domain data for the test, and the remaining 80\% of new domain data are the training data.
We use different portions (1\%, 25\%, 50\% 75\%, 100\%) of training data to fine-tune the EnsembleDLM and compare its performance with DeepDTA \cite{ozturk2018deepdta} and DeepCDA \cite{abbasi2020deepcda}.
We implement five-fold cross-validation on the training set.
The results are presented in Table \ref{tab:tf_davis_kiba}. 

\begin{table*}[htbp!]
\renewcommand*{\arraystretch}{1.4}
\centering
\caption{The cross-domain performance comparisons of our approach and state-of-the-art approaches on Davis and KIBA. The results are rounded down to two decimal places.}
\resizebox{\textwidth}{!}{%
\begin{tabular}{@{\extracolsep{4pt}}lcccccccc@{}}
\toprule
& \multicolumn{4}{c}{Davis $\rightarrow$ KIBA} & \multicolumn{4}{c}{KIBA $\rightarrow$ Davis} \\ \cline{2-5}\cline{6-9}
& \multicolumn{2}{c}{Validation} & \multicolumn{2}{c}{Test} & \multicolumn{2}{c}{Validation} & \multicolumn{2}{c}{Test} \\ \cline{2-3}\cline{4-5}\cline{6-7}\cline{8-9}
& C-index$\pm$std & PCC$\pm$std  & C-index$\pm$std & PCC$\pm$std & C-index$\pm$std & PCC$\pm$std & C-index$\pm$std & PCC$\pm$std \\ 
\midrule
DeepDTA\cite{ozturk2018deepdta} & 0.55$\pm$0.01 & - & - & - & 0.67$\pm$0.04 & - & - & - \\
DeepCDA\cite{abbasi2020deepcda} & 0.58$\pm$0.01 & - & - & - & 0.68$\pm$0.03 & - & - & - \\
Proposed Approach & \textbf{0.59$\pm$0.01} & 0.32$\pm$0.01 & 0.59$\pm$0.00 & 0.32$\pm$0.01 & \textbf{0.73$\pm$0.01} & 0.52$\pm$0.02 & 0.73$\pm$0.01 & 0.52$\pm$0.01 \\ \hline \hline
DeepDTA\cite{ozturk2018deepdta} with DA & 0.60$\pm$0.04 & - & - & - & 0.72$\pm$0.04 & - & - & - \\
DeepCDA\cite{abbasi2020deepcda} with DA & 0.64$\pm$0.03 & - & - & - & 0.74$\pm$0.03 & - & - & - \\
Proposed Approach with TL (1\%)  & 0.71$\pm$0.01 & 0.53$\pm$0.01 & 0.71$\pm$0.01 & 0.53$\pm$0.01 & 0.75$\pm$0.01 & 0.53$\pm$0.02 & 0.76$\pm$0.01 & 0.54$\pm$0.01 \\
Proposed Approach with TL (25\%) & 0.85$\pm$0.00 & 0.84$\pm$0.00 & 0.85$\pm$0.00 & 0.84$\pm$0.00 & 0.87$\pm$0.01 & 0.79$\pm$0.00 & 0.87$\pm$0.00 & 0.78$\pm$0.01 \\
Proposed Approach with TL (50\%) & 0.87$\pm$0.00 & 0.87$\pm$0.00 & 0.87$\pm$0.00 & 0.87$\pm$0.00 & 0.89$\pm$0.00 & 0.84$\pm$0.01 & 0.89$\pm$0.00 & 0.83$\pm$0.00 \\
Proposed Approach with TL (75\%) & 0.88$\pm$0.00 & 0.89$\pm$0.00 & 0.88$\pm$0.00 & \textbf{0.89$\pm$0.00} & 0.90$\pm$0.00 & 0.86$\pm$0.01 & \textbf{0.90$\pm$0.00} & 0.85$\pm$0.00 \\
Proposed Approach with TL (100\%)& \textbf{0.89$\pm$0.00} & \textbf{0.90$\pm$0.00} & \textbf{0.89$\pm$0.00} & \textbf{0.89$\pm$0.00} & \textbf{0.91$\pm$0.00} & \textbf{0.87$\pm$0.01} & \textbf{0.90$\pm$0.00} & \textbf{0.86$\pm$0.00} \\
\bottomrule
\end{tabular}
}
\begin{tablenotes}
\item DA: domain adaptation. TL: transfer learning. 
\end{tablenotes}
\label{tab:tf_davis_kiba}
\end{table*}

\subsubsection{Different protein classes with same type of binding affinity scores}
We then evaluate the performance of EnsembleDLM across different protein classes with the same binding affinity score.
The ChEMBL-REST dataset is used as the original domain, and the EnsembleDLM is fine-tuned to new domains (ChEMBL-ER, ChEMBL-TFNR, ChEMBL-EK, and ChEMBL-EP).
These datasets only contain the $K_i$ drug-target interactions.
The protein classes in the ChEMBL-ER, ChEMBL-TFNR, ChEMBL-EK and ChEMBL-EP sets are excluded in the ChEMBL-REST set.
We first divide 20\% of new domain data for test and 80\% of new domain data for training. 
We then use different portions (1\%, 25\%, 50\%, 75\%, 100\%) of new domain training data to fine-tune the base EnsembleDLM with five-fold cross-validation.
The results are presented in Table \ref{tab:tf_chembl}.

\begin{table*}[htbp!]
\renewcommand*{\arraystretch}{1.4}
\centering
\caption{The cross-domain performance of our approach on $K_i$ interactions of ChEMBL dataset between different protein classes with 5-fold cross-validation. Different sizes of new domain data are used to fine-tune the base EnsembleDLM.  The results are rounded down to two decimal places, and the bold number highlights the best performance.}
\begin{tabular}{@{\extracolsep{4pt}}ccccccccc@{}}
\toprule
 & \multicolumn{8}{c}{ChEMBL} \\ \cmidrule{2-9}
 & \multicolumn{2}{c}{REST $\rightarrow$ ER} & \multicolumn{2}{c}{REST $\rightarrow$ TFNR} & \multicolumn{2}{c}{REST $\rightarrow$ EK}    & \multicolumn{2}{c}{REST $\rightarrow$ EP} \\ \cmidrule{2-3} \cmidrule{4-5} \cmidrule{6-7} \cmidrule{8-9}  
Size & PCC$\pm$std  & C-index$\pm$std & PCC$\pm$std & C-index$\pm$std & PCC$\pm$std  & C-index$\pm$std & PCC$\pm$std & C-index$\pm$std \\ \midrule \midrule
0\%     & 0.148$\pm$0.081 & 0.550$\pm$0.027 &-0.016$\pm$0.051 & 0.504$\pm$0.016 & 0.263$\pm$0.022 & 0.589$\pm$0.008 & 0.225$\pm$0.026 & 0.578$\pm$0.007 \\
1\%     & 0.391$\pm$0.074 & 0.638$\pm$0.024 & 0.304$\pm$0.059 & 0.601$\pm$0.016 & 0.608$\pm$0.022 & 0.714$\pm$0.009 & 0.442$\pm$0.023 & 0.656$\pm$0.006 \\
25\%    & 0.776$\pm$0.026 & 0.799$\pm$0.013 & 0.774$\pm$0.014 & 0.789$\pm$0.007 & 0.843$\pm$0.005 & 0.827$\pm$0.001 & 0.836$\pm$0.006 & 0.829$\pm$0.003 \\
50\%    & 0.828$\pm$0.030 & 0.828$\pm$0.016 & 0.824$\pm$0.011 & 0.817$\pm$0.006 & 0.871$\pm$0.005 & 0.846$\pm$0.001 & 0.866$\pm$0.002 & 0.850$\pm$0.003 \\
75\%    & 0.849$\pm$0.019 & 0.843$\pm$0.008 & 0.846$\pm$0.008 & 0.831$\pm$0.004 & 0.885$\pm$0.005 & 0.857$\pm$0.001 & 0.878$\pm$0.002 & 0.861$\pm$0.003 \\ 
100\%   & \textbf{0.860$\pm$0.017} & \textbf{0.849$\pm$0.008} & \textbf{0.854$\pm$0.006} & \textbf{0.836$\pm$0.004} & \textbf{0.891$\pm$0.004} & \textbf{0.862$\pm$0.001} & \textbf{0.883$\pm$0.003} & \textbf{0.866$\pm$0.002} \\\bottomrule
\end{tabular}
\label{tab:tf_chembl}
\end{table*}

\subsubsection{Different protein classes with different types of binding affinity scores}

In this experiment, we would like to examine the cross-domain performance of EnsembleDLM across different protein classes with different binding affinity scores.
ChEMBL-REST is used as the base (original) domain.
It contains only $K_i$ interaction of drug-target pairs, and the interactions of the kinase are not included in this set. 
The KIBA and Davis datasets are used as new domains.
KIBA comprises \textit{KIBA} scores from the interactions of kinase, and the KIBA score is integrated by $IC_{50}$, $K_d$, and $K_i$.  
Davis consists of $K_d$ interactions of drugs and kinase.
We left 20\% of new domain data for the test, and 80\% of the remaining new domain data are the training data.
EnsembleDLM is first trained by interactions of the base domain (ChEMBL-REST) and then fine-tuned to the new domain (KIBA, Davis) by using different portions (1\%, 25\%, 50\%, 75\%, 100\%) of new domain training interactions.
The results are shown in Table \ref{tab:tf_chembl_davis_kiba}

\begin{table*}[htbp!]
\renewcommand*{\arraystretch}{1.4}
\centering
\caption{The cross-domain performance comparisons of EnsembleDLM. The EnsembleDLM is first trained by whole data of ChEMBL-REST set, and then fine-tuned to Davis and KIBA set using different portions of new domain data. The results are rounded down to two decimal places.}
\resizebox{\textwidth}{!}{%
\begin{tabular}{@{\extracolsep{4pt}}lcccccccc@{}}
\toprule
& \multicolumn{4}{c}{ChEMBL-REST $\rightarrow$ KIBA} & \multicolumn{4}{c}{ChEMBL-REST $\rightarrow$ Davis} \\ \cline{2-5}\cline{6-9}
& \multicolumn{2}{c}{Validation} & \multicolumn{2}{c}{Test} & \multicolumn{2}{c}{Validation} & \multicolumn{2}{c}{Test} \\ \cline{2-3}\cline{4-5}\cline{6-7}\cline{8-9}
& C-index$\pm$std & PCC$\pm$std  & C-index$\pm$std & PCC$\pm$std & C-index$\pm$std & PCC$\pm$std & C-index$\pm$std & PCC$\pm$std \\ 
\midrule
Proposed Approach without TL & 0.54$\pm$0.01 & 0.12$\pm$0.03 & 0.54$\pm$0.01 & 0.12$\pm$0.02  & 0.51$\pm$0.02 & 0.03$\pm$0.02 & 0.52$\pm$0.02 & 0.04$\pm$0.003 \\
Proposed Approach with TL (1\%)  & 0.67$\pm$0.00 & 0.47$\pm$0.01 & 0.67$\pm$0.00 & 0.46$\pm$0.01 & 0.72$\pm$0.02 & 0.43$\pm$0.02 & 0.73$\pm$0.01 & 0.44$\pm$0.02 \\
Proposed Approach with TL (25\%) & 0.84$\pm$0.00 & 0.82$\pm$0.00 & 0.84$\pm$0.00 & 0.82$\pm$0.00 & 0.86$\pm$0.01 & 0.77$\pm$0.01 & 0.86$\pm$0.00 & 0.76$\pm$0.00 \\
Proposed Approach with TL (50\%) & 0.88$\pm$0.00 & 0.87$\pm$0.00 & 0.88$\pm$0.00 & 0.87$\pm$0.00 & 0.89$\pm$0.00 & 0.82$\pm$0.01 & 0.89$\pm$0.00 & 0.81$\pm$0.01 \\
Proposed Approach with TL (75\%) & 0.89$\pm$0.00 & 0.89$\pm$0.00 & 0.89$\pm$0.00 & 0.89$\pm$0.00 & 0.90$\pm$0.01 & 0.85$\pm$0.01 & \textbf{0.90$\pm$0.00} & 0.84$\pm$0.00 \\
Proposed Approach with TL (100\%)& \textbf{0.90$\pm$0.00} & \textbf{0.90$\pm$0.00} & \textbf{0.90$\pm$0.00} & \textbf{0.90$\pm$0.00} & \textbf{0.91$\pm$0.00} & \textbf{0.86$\pm$0.01} & \textbf{0.90$\pm$0.00} & \textbf{0.85$\pm$0.00} \\
\bottomrule
\end{tabular}
}
\begin{tablenotes}
\item TL: transfer learning. 
\end{tablenotes}
\label{tab:tf_chembl_davis_kiba}
\end{table*}

\section{Discussion and Conclusion}

We first demonstrate that EnsembleDLM has better performance than any single model within EnsembleDLM and the sequence-based state-of-the-art methods. 
From Table \ref{tab:single_model_davis_kiba} and \ref{tab:single_model_chembl}, EnsembleDLM has a remarkable improvement compared to any single model within the EnsembleDLM in Davis, KIBA, and ChEMBL-REST datasets.
It also has the smallest standard deviations compared to other single models. 
Compared to the best model (Morgan-CNN) within the EnsembleDLM, EnsembleDLM has approximately 17.21\% and 2.95\% improvements on the MSE and CI in Davis validation sets, and 12.82\% and 2.72\% improvements on the MSE and CI in Davis test set, respectively.
It also has approximately 19.3\% and 1.7\% improvements on the MSE and CI in KIBA validation sets, and 18.34\% and 1.7\% improvements on the MSE and CI in the KIBA test set, respectively, compared to the best model (Morgan-CNN) within the EnsembleDLM.
Moreover, EnsembleDLM has approximately 25\% and 2.8\% improvements on the MSE and CI in ChEMBL-REST validation sets, and 21.72\% 2.67\% improvements on the MSE and CI in ChEMBL-REST test set, respectively, compared to the best model (Morgan-AAC) within the EnsembleDLM. 
From Table \ref{tab:davis} and \ref{tab:kiba}, EnsembleDLM has the best performance compared to the state-of-the-art methods in both Davis and KIBA datasets.
It is noted that we only compare to the sequence-based methods for a fair comparison. 

To evaluate the cross-domain performance of EnsembleDLM, we come up with three different scenarios: 1.) same protein class with different bio-activity types, 2.) different protein classes with the same bio-activity type, and 3.) different protein classes with different bio-activity types. 
Davis and KIBA datasets are used for the case of the same protein class with different bio-activity types.
They comprise kinase inhibitors bio-activity data but with different bio-activity types ($K_d$ for Davis and \textit{KIBA} for KIBA).
Davis and KIBA datasets have 9 overlapping drugs and 179 overlapping targets.
From Table \ref{tab:tf_davis_kiba}, EnsembleDLM has better cross-domain drug-target interaction prediction performance compared to the state-of-the-art methods without transfer learning or domain adaptation in both Davis $\rightarrow$ KIBA and KIBA $\rightarrow$ Davis. 
With only 1\% transfer learning data, EnsembleDLM has better performance than the state-of-the-art methods with domain adaptation in both Davis $\rightarrow$ KIBA and KIBA $\rightarrow$ Davis.
With 100\% transfer learning data, EnsembleDLM approaches similar performance compared to the training-from-scratch EnsembleDLM shown in Table \ref{tab:single_model_davis_kiba}. 
Bayesian optimization is not applied for hyper-parameter searching in the transfer learning phase.
Therefore, the performance of the training-from-scratch EnsembleDLM is marginally better than the performance of the transfer-learning EnsembleDLM. 
Then, ChEMBL datasets are used to evaluate the cross-domain performance of EnsembleDLM for the case of different protein classes with the same bio-activity type.
Table \ref{tab:tf_chembl} shows that without the transfer learning, EnsembleDLM has the best performance on the EK dataset and the worst performance on the TFNR dataset. 
With approximately 25\% transfer learning data, EnsembleDLM has a good (PCC and C-index $>$ 0.8) performance in EK and EP datasets.
EnsembleDLM also has a good (PCC and C-index $>$ 0.8) performance in ER and TFNR datasets with approximately 50\% transfer learning data.
With 100\% transfer learning data, the EnsembleDLM again has the best performance on the EK dataset followed by EP, ER, and TFNR dataset. 
In the last scenario, we would like to evaluate the cross-domain performance of our model across different protein families with different bio-activity types.  
ChEMBL-REST is used as the base (original) dataset, and KIBA and Davis are used as new domain datasets. 
KIBA and Davis contain interactions of kinases that are not curated in the ChEMBL-REST set. 
The bio-activity type of ChEMBL-REST is $K_i$, whereas the bio-activity type is $K_d$ and \textit{KIBA} for the Davis and KIBA dataset, respectively. 
Different portions (1\%, 25\%, 50\%, 75\%, and 100\%) of data from new domains are used for transfer learning. 
Table \ref{tab:tf_chembl_davis_kiba} shows that without transfer learning, EnsembleDLM has better performance in ChEMBL-REST $\rightarrow$ KIBA.
With approximately 25\% transfer learning data, EnsembleDLM has a good (PCC and C-index $>$ 0.8) performance in both ChEMBL-REST $\rightarrow$ KIBA and ChEMBL-REST $\rightarrow$ Davis. 
With 100\% transfer learning data, EnsembleDLM has similar performance no matter the base (original) datasets are from the same protein class (Davis $\rightarrow$ KIBA and KIBA $\rightarrow$ Davis) or the different protein classes (ChEMBL-REST $\rightarrow$ KIBA and ChEMBL-REST $\rightarrow$ Davis).
However, for different protein families analysis, EnsembleDLM has better performance for the same bio-activity type (ChEMBL-REST $\rightarrow$ ChEMBL-EK) than different bio-activity types (ChEMBL-REST $\rightarrow$ KIBA and ChEMBL-REST $\rightarrow$ Davis).

In conclusion, EnsembleDLM achieves better performance compared to any single model within the EnsembleDLM and the sequence-based state-of-the-art methods. 
It also has better cross-domain performance compared to the sequence-based state-of-the-art methods. 
With approximately 50\% transfer learning data, i.e., the training set has twice as much data as the test set, EnsembleDLM achieves a good performance (PCC and C-index $>$ 0.8) in the new domain in every scenario.
However, the EnsembleDLM only has marginal improvements from 50\% to 75\% of transfer learning data and from 75\% to 100\% of transfer learning data.

\section*{Acknowledgment}

We would like to thank Kexin Huang and Tianfan Fu for technical support on DeepPurpose, Dr. Karim Abbasi for his assistance on DeepCDA, and Taiwan Computing Cloud (TWCC) for computational resources. 

\bibliographystyle{./bibliography/IEEEtran}
\bibliography{./bibliography/IEEE}

\end{document}